\documentclass[12pt,journal,compsoc,onecolumn]{IEEEtran}
\usepackage{booktabs} % For formal tables
\usepackage[ruled]{algorithm2e} % For algorithms
\usepackage{array} 
\usepackage{wrapfig}
\usepackage{graphicx}
\usepackage{longtable}
\usepackage{lineno}
\usepackage{framed}
\usepackage{float}
\restylefloat{table}
\usepackage{rotating}
\usepackage{subfigure}
\usepackage{enumitem}
\usepackage{color}
\usepackage{bbding}

\usepackage{wrapfig}
\usepackage{longtable}
\usepackage{lineno}
\usepackage{framed}
\usepackage{float}
\restylefloat{table}
\usepackage{rotating}
\usepackage{enumitem}
\usepackage{color}
\usepackage{framed}
\usepackage{colortbl}
\usepackage{amsmath}

\usepackage{tabularx}
\usepackage{url}
\usepackage{xcolor}
\usepackage{bbding}
\usepackage{url}
\usepackage{balance}
\usepackage{multirow}
\usepackage{booktabs}
\usepackage{graphicx}
\usepackage[framemethod=tikz]{mdframed}

\usepackage{amsmath}
\usepackage{amssymb}
\usepackage{graphicx}
\usepackage{rotating}

\usepackage{amsmath}

\usepackage{algorithmic}

\usepackage{array}
\usepackage{multirow}% http://ctan.org/pkg/multirow
\usepackage{hhline}% http://ctan.org/pkg/hhline
\usepackage[ruled]{algorithm2e} % For algorithms
\usepackage{array} 
\usepackage{wrapfig}
\usepackage{graphicx}
\usepackage{longtable}
\usepackage{lineno}
\usepackage{framed}
\usepackage{float}
\restylefloat{table}
\usepackage{rotating}
\usepackage{subfigure}
\usepackage{enumitem}
\usepackage{color}
\usepackage{bbding}

\usepackage{wrapfig}
\usepackage{longtable}
\usepackage{lineno}
\usepackage{framed}
\usepackage{float}
\restylefloat{table}
\usepackage{rotating}
\usepackage{enumitem}
\usepackage{color}
\usepackage{framed}
\usepackage{colortbl}
\usepackage{amsmath}

\usepackage{tabularx}
\usepackage{url}
\usepackage{xcolor}
\usepackage{bbding}
\usepackage{url}
\usepackage{balance}
\usepackage{multirow}
\usepackage{booktabs}
\usepackage{graphicx}
\usepackage[framemethod=tikz]{mdframed}

\usepackage{amsmath}
\usepackage{amssymb}
\usepackage{graphicx}
\usepackage{rotating}

\usepackage[ruled]{algorithm2e}

\usepackage{array} 
\usepackage{wrapfig}
\usepackage{graphicx}
\usepackage{longtable}
\usepackage{lineno}
\usepackage{framed}
\usepackage{float}
\restylefloat{table}
\usepackage{rotating}
\usepackage{subfigure}
\usepackage{enumitem}
\usepackage{color}
\usepackage{framed}
\usepackage{colortbl}
\usepackage{amsmath}
\usepackage{subfigure}
\usepackage{url}
\usepackage{xcolor}
\usepackage{bbding}
\usepackage{url}

\usepackage{amsmath}

\usepackage{algorithmic}

\newmdenv[innerlinewidth=0.5pt, roundcorner=4pt,innerleftmargin=6pt,
innerrightmargin=6pt,innertopmargin=6pt,innerbottommargin=6pt]{mybox}

{

\usepackage{array}
\usepackage{multirow}% http://ctan.org/pkg/multirow
\usepackage{hhline}% http://ctan.org/pkg/hhline
\pagestyle{plain}
\newcommand\revised[1]{\textcolor{black}{#1}}

\begin{document}
\hyphenation{op-tical net-works semi-conduc-tor}
% Titles are generally capitalised except for words such as a, an, and, as,
% not capitalised unless they are the first or last word of the title.
%\title{Organizational Structure Patterns in Agile Software Teams: An Industrial Study}
%\IEEEcompsocitemisethanks is a special \thanks that produces the bulleted
% \IEEEcompsocitemisethanks becomes like \thanks and
% one-sise-fits-all approach a daunting prospect. For instance, compsoc 
\author{Damian A.~Tamburri,~\IEEEmembership{Member-at-Large,~IEEE,}
Fabio~Palomba,~\IEEEmembership{Member~IEEE,}
        Rick Kazman,~\IEEEmembership{Senior Member,~IEEE}\\
        
        % $<$-this % stops a space
\IEEEcompsocitemizethanks{\IEEEcompsocthanksitem D. A. Tamburri is with the Eindhoven University of Technology and the Jheronimus academy of Data Science\protect\\
E-mail: d.a.tamburri@tue.nl

\IEEEcompsocthanksitem F. Palomba is with University of Salerno
E-mail: fpalomba@unisa.it

\IEEEcompsocthanksitem R. Kazman is with University of Hawaii \& SEI/CMU
E-mail: kazman@hawaii.edu}}
\title{Success and Failure in Software Engineering:\\ a Followup Systematic Literature Review}
 %\titlenote{This is a titlenote}
 %\subtitle{This is a subtitle}
 %\subtitlenote{Subtitle note}

%\author{Damian A. Tamburri}
%\affiliation{%
%  \institution{Eindhoven University of Technology}
%  \streetaddress{Sint janssingel 92}
%  \city{s'Hertogenbosch}
%  \state{NB}
%  \postcode{5711MN}
%  \country{NL}}
%\email{d.a.tamburri@tue.nl}
%\author{Fabio Palomba}
%\affiliation{%
% \institution{University of Zurich}
% \city{Zurich}
% \country{Switzerland}}
%\email{palomba@ifi.uzh.ch}
%\author{Rick Kazman}
%\affiliation{%
%  \institution{University of Hawaii / SEI-CMU}
%  \city{Manoa}
%  \country{US}
%}
%\email{kazman@hawaii.edu}

%\renewcommand\shortauthors{Zhou, G. et al}
\IEEEtitleabstractindextext{%
\begin{abstract}
%Software engineering is increasingly a community effort and its success depends on balancing distance, culture, global engineering practices. However there are many socio-technical factors that may increase project cost or ``social" debt. Employee or contributor dissatisfaction has reportedly led to higher turnover, reduced effort or even ``rage-quitting", that is, when people abandon an open- or closed-source software community slamming the door.  These factors  require attention from community leaders---architects, project managers, and core committers in open-source communities---to mitigate the effects connected to social anti-patterns and negative characteristics. To raise the awareness of community issues as well as to further understand and support a software development community in improving and making its endeavors sustainable, we have conducted empirical research to elicit, operationalize, and evaluate a \emph{community quality model}: a set of metrics to track organizational and socio-technical qualities that are key to managing social debt. This article reports on the quality model starting from a thorough detail of the systematic literature study that led to its inception.
\revised{Success and failure in software engineering are still among the least understood phenomena in the discipline. In a recent special journal issue on the topic, M{\"a}ntyl{\"a} et al. started discussing these topics from different angles; the authors focused their contributions on offering a general overview of both topics without deeper detail. Recognising the importance and impact of the topic, we have executed a followup, more in-depth systematic literature review with additional analyses beyond what was previously provided. These new analyses offer: (a) a grounded-theory of success and failure factors, harvesting over 500+ factors from the literature; (b) 14 manually-validated clusters of factors that provide relevant areas for success- and failure-specific measurement and risk-analysis; (c) a quality model composed of previously unmeasured organizational structure quantities which are germane to software product, process, and community quality. We show that the topics of success and failure deserve further study as well as further automated tool support, e.g.,  monitoring tools and metrics able to track the factors and patterns emerging from our study. This paper provides managers with risks as well as a more fine-grained analysis of the parameters that can be appraised to anticipate the risks.}
\end{abstract}

\begin{IEEEkeywords}
Success and failure; Software engineering; Systematic Literature Reviews.
\end{IEEEkeywords}}

%\begin{CCSXML}
%<ccs2012>
%<concept>
%<concept_id>10002944.10011123.10010912</concept_id>
%<concept_desc>General and reference~Empirical studies</concept_desc>
%<concept_significance>500</concept_significance>
%</concept>
%<concept>
%<concept_id>10003120.10003130.10011762</concept_id>
%<concept_desc>Human-centered computing~Empirical studies in collaborative and %social computing</concept_desc>
%<concept_significance>300</concept_significance>
%</concept>
%<concept>
%<concept_id>10003456.10003462</concept_id>
%<concept_desc>Social and professional topics~Computing / technology policy</concept_desc>
%<concept_significance>100</concept_significance>
%</concept>
%</ccs2012>
%\end{CCSXML}

%\ccsdesc[500]{General and reference~Empirical studies}
%\ccsdesc[300]{Human-centered computing~Empirical studies in collaborative and social computing}
%\ccsdesc[100]{Social and professional topics~Computing / technology policy}
\maketitle
\IEEEdisplaynontitleabstractindextext
\IEEEpeerreviewmaketitle
%\keywords{Organizational Structures, Human Aspects in Software Engineering, Systematic Literature Review}

%\maketitle

\section{Introduction}
In the scope of software production and operation, the notions of success and failure are intriguing, having different forms and manifesting under varied conditions \cite{RalphK14,LehtinenMVIL14}. In a recent special issue of {\em Empirical Software Engineering} on this topic \cite{MantylaJRE17}, the editors remarked that, \emph{``despite ongoing concerns over the failure rate of software projects,  basic questions such as ``How do we measure general software success?" and ``How can software failure rates be measurably reduced?" remain still only partially explored"}. The editors concluded that addressing these questions is critical to further understand and steer software projects towards success. We pick up the challenge from where it was left off \cite{MantylaJRE17}. In this paper we refine and re-execute the research design set up by the editors in their special issue introduction, aimed at identifying and analyzing a set of papers focused on the topics of success and failure in software engineering research and practice.

The goal we address is to add further analyses on top of what M{\"a}ntyl{\"a} et al. \cite{MantylaJRE17} offer as a preliminary analysis. 
Our objectives with these additional analyses are threefold. First, we aim to elicit a {\em grounded theory} of success and failure factors so that other researchers may identify such factors and how to measure them, ideally creating a general software success (or failure) prediction model. Second, we aim to highlight the most relevant themes of factors thus identifying the areas of software engineering research and practice that are under-supported by measures. Third, we aim to elicit a rigorous quality model for these under-supported quantities.

Briefly, our results show that success and failure in software projects is mediated by over 500 factors (e.g., presence of users directly in the software process \cite{iivari_enculturation_2004}) arranged in 40+ core-concepts (e.g., effort estimation). Furthermore, there exist 14 themes along which success and failure is determined (in practice) and studied (in research), such as \emph{best practices evaluation and monitoring}, or \emph{software measurement} or \emph{organizational structure and motivation}. Finally, 5 out of the 14  themes reflect organizational structure quality which, to date, does not have a rigorous model (that is, a set of measurable quantities \cite{standard2003a}). \revised{As the final contribution of this article, we offer a first attempt at such a quality model that captures the most recurrent measurable factors and quantities from the aforementioned 5 themes, such as  \emph{truck-factor} \cite{TorchianoRM11} or \emph{socio-technical congruence} \cite{sociotechnical}.}\\

\noindent \revised{\textbf{Replication Package.} Finally, to encourage replication we make available a comprehensive package containing all papers, Grounded-Theory sources as well as analysis of data performed in this study~\footnote{See here: \url{https://figshare.com/s/e6f0968e55c2cd024389}}.\\}

\noindent \revised{\textbf{Structure of the paper.}
Section \ref{terminology} outlines the terminology used in the paper. In Section \ref{mm}, we describe our research methods, while Section \ref{res} overviews the results achieved. Section \ref{disc} provides discussions on the key findings of the paper. Finally, Section \ref{conc} concludes the paper and outlines our future research agenda on the topic.} 

%\subsection{Terminology}

%Because of the nature of a followup study, we inherit much of the terminology previously introduced or defined in the target study \cite{MantylaJRE17}. In that respect, the \emph{context} of our study reflects successful or failed software engineering projects in industry or their study from any empirical, experimental, theoretical perspective. The concepts of \emph{success} and \emph{failure} are defined, respectively, as (1) the long-lasting conditions wherefore a software is maintained fitting with its expectations and (2) the moment in time and software operation wherefore the aforementioned conditions stop existing.

\section{Scope and Terminology}
\label{terminology}
\revised{
The scope intended for this work draws primarily from the single preliminary study reported in M\"{a}ntyl\"{a} et al. \cite{MantylaJRE17} which encompasses a very large sample of research and discusses the concepts of success and failure or the context in which such phenomena manifest themselves from a very high-level. The scope we set out to investigate as a spin-off of the aforementioned previous work encompasses the high-relevance and high-impact research currently available in literature that elaborates either on (1) the primary studies emerged in M\"{a}ntyl\"{a} et al. \cite{MantylaJRE17} or (2) on any of the concepts or conclusions emerged  in the same paper. With respect to point 1 above, We are aware that these phenomena are complex and cannot be simplistically reduced to mere factors and dimensions. Our goal is to build upon the work by M\"{a}ntyl\"{a} et al. and consequently collect a grounded-theory which acts as a foundation of what is  known about these phenomena such that further work can be developed based on this foundation. With respect to point 2 above, our paper is a followup study to M\"{a}ntyl\"{a} et al. and, for this reason, we inherit much of the terminology previously used in the target study \cite{MantylaJRE17}. In the following, we report those terms and their associated meaning:}

\begin{description}[leftmargin=0.3cm]
	\smallskip
	\item[Context.] \revised{This reflects successful or failed software engineering projects and their study from any empirical, experimental, or theoretical perspective.}
	
	\smallskip
	\item[Success.] \revised{Success represents the long-lasting conditions wherefore a software project is maintained in a state meeting its expectations.}
	
	\smallskip
	\item[Failure.] \revised{A failure is the moment in time where a software project no longer meets its expectations.}
	
\end{description}

\revised{In the next section, we describe the research methodology we employed to conduct our followup systematic literature review.}

\section{Research Approach}\label{mm}
\subsection{Research Design}
The \emph{goal} of this work is to obtain an in-depth overview on the phenomena of software success and failure. The \emph{purpose} is to provide the research community with actionable insights on the factors impacting a software project to be successful, so that future studies could be devised to explicitly target novel methodologies and methods to take those factors under control. Our \emph{perspective} is of both researchers and practitioners, who are interested in gathering deeper knowledge of the attributes to be monitored to mitigate the risk of software failure. 

To this end we aimed at providing further analyses on top of the literature retrieved to provide a greater depth of understanding. The analyses we added aim at answering the following \emph{research questions} (\textbf{RQ}s):

\begin{enumerate}
\item[\textbf{RQ$_1$}.] \emph{What factors are reportedly connected to success or failure?}
\item[\textbf{RQ$_2$}.] \emph{What themes emerge across such factors?}
\item[\textbf{RQ$_3$}.] \emph{What themes are currently unobserved and what previously-existing metrics can support these unobserved themes?}
\end{enumerate}

\begin{figure}
\begin{center}
\includegraphics[scale=0.5]{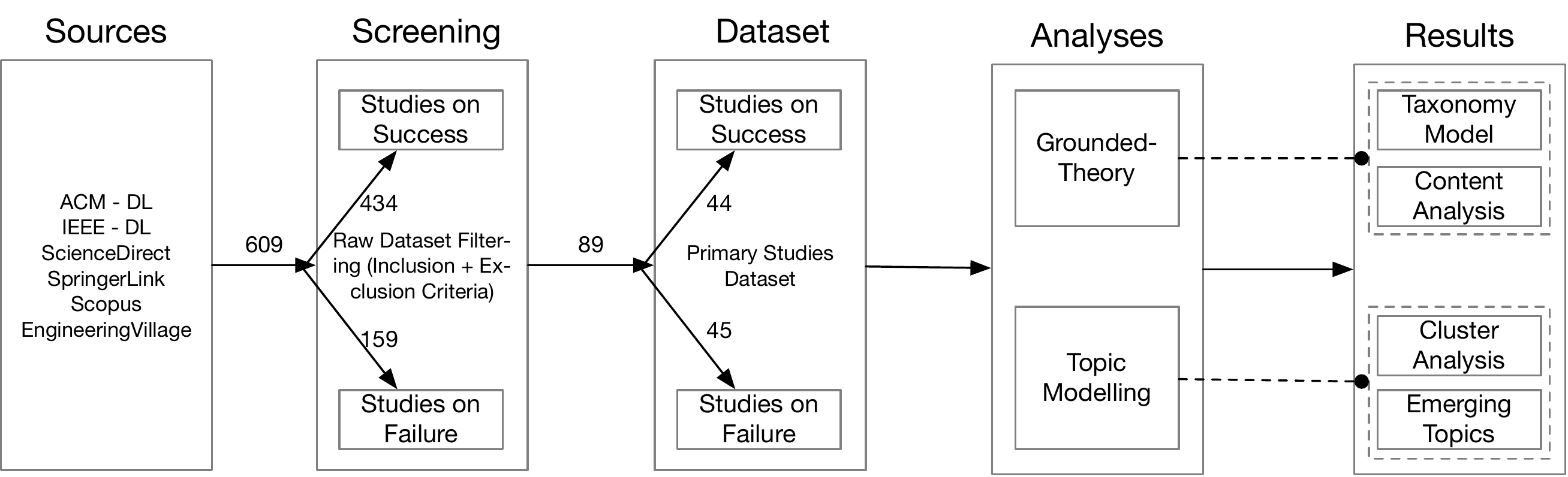}
\end{center}
\caption{\revised{An overview of our research design, from sources to results.}}\label{methodoverview}
\end{figure}

Note that, in the scope of \textbf{RQ$_3$}, by \emph{unobserved} we mean the factors and themes emerging from \textbf{RQ$_1$} and \textbf{RQ$_2$} that currently have no accepted metrics to support their appraisal. The ultimate goal of this research question is to provide practitioners with a \emph{quality model}, that is, an aggregate of measures which were previously defined, evaluated, and automated for these unobserved factors and themes. \revised{Fig. \ref{methodoverview} recaps the main steps undertaken to attain results as well as the inputs and outputs of each phase using a simple box and line diagram. The main boxes in the figure represent steps that were undertaken while smaller boxes represent results of those steps linked by action arrows; for the early dataset and sample selection stages,  arrows are augmented with quantities connected to the sampling process. Finally, dotted lines connect each analysis (quantitative or qualitative) with its analysis results.}

%\revised{More in general, we address our research questions by means of a followup literature review. We adopted this method because a systematic review of the literature can let emerge a set of factors that previous researchers have empirically verified to be connected to success and failure of software engineering projects, thus allowing us to provide  comprehensive framework of measures that project managers could keep under control when assessing the status of their projects. In the following, we describe the specific research approach conducted.}

\subsection{Literature Retrieval Approach}
%The systematic literature review approach is generally organized in three sequential steps: \textbf{(a)} Create the set of articles to review (i.e. primary studies); \textbf{(b)} Conduct the review; \textbf{(c)} Analyze collected data.

To retrieve the target literature we executed an augmented  retrieval strategy described in previous work \cite{MantylaJRE17}. The new strategy focuses on eliciting papers that focus on industrial applicability of the proposed claims, results, and contributions or which offer results stemming from industrial practice and experience. Specifically, to retrieve papers we execute the following search string:

\medskip
\begin{center}
	\fbox{
		\begin{minipage}[t]{0.9\linewidth}
			TITLE-ABS ((``software engineering" OR ``software development" OR ``software project" OR ``it project" OR ``it development" OR ``it engineering") AND TITLE (``success" OR ``failure") AND BODY (``case-study" OR ``industrial-*" OR ``practiction*"))
		\end{minipage}
	}
\end{center}
\medskip

where TITLE-ABS indicates that the subsequent search terms are considered only in the scope of title and abstract of the papers.  TITLE indicates that the search is conducted only on papers titles whereas BODY indicates that the search is conducted only on the body of the articles. This is the exact search string defined by M{\"a}ntyl{\"a} et al. %\cite{MantylaJRE17} after a number of iterations aimed at establishing the most suitable set of keywords to use. For instance, they found that several articles not related to either success or failure in software engineering used the words ``success" and ``failure" to motivate their work; for this reason, they required that the terms ``success" and ``failure" appear only in the paper title. At the same time, when considering papers about failure, the authors realized that many false positives were due to the presence of papers focusing on failure engineering and reliability modelling that consider software engineering failures only as part of software that is malfunctioning. This would have created inconsistencies with respect to the focus on entire software projects and processes.

%For this reason, these papers were excluded. To automate the removal process, the authors of our reference study defined a search string that, if matched, led to the exclusion of a paper: 

%\medskip
%\begin{center}
%	\fbox{
%		\begin{minipage}[t]{0.9\linewidth}
%			TITLE-ABS-KEY (``heart failure" OR ``failure mode" OR ``stability failure" OR ``failure mode" OR ``ceramic" OR ``steel" OR ``mtbe" OR ``coal" OR ``fault diagnosis" OR ``system failure engineering" OR ``Failure simulation" OR ``slope failure" OR ``fault tolerance" OR ``physiology" OR ``reliability modeling" OR ``software reliability")
%		\end{minipage}
%	}
%\end{center}
%\medskip

%where TITLE-ABS-KEY indicates that the terms are verified in title, abstract, and keywords of the papers. Following this string, papers reporting any of the above mentioned terms were automatically excluded from the final list of primary studies. 
\revised{The search process has been conducted on a number of different databases, namely:}

	\begin{itemize}
	    \item \revised{IEEE Xplore digital library (http://ieeexplore.ieee.org);}
		\item \revised{ACM digital library (https://dl.acm.org);}
		\item \revised{ScienceDirect (http://www.sciencedirect.com);}
		\item \revised{SpringerLink (https://link.springer.com);}
        \item \revised{Scopus (https://www.scopus.com);}
        \item \revised{Engineering Village (https://www.engineeringvillage.com);}
	\end{itemize}

\revised{The selection of these databases was driven by our willingness to gather as many papers as possible to properly conduct our systematic literature review. In this respect, the selected sources are recognized as the most representative and complete for Software Engineering research and are used in many other SLRs \cite{Kitchenham2007} because they contain a massive amount of literature---journal articles, conference proceedings, books etc.---related to our research questions. As described by M{\"a}ntyl{\"a} et al. \cite{MantylaJRE17}, no paper on success and failure in software engineering was published before 1970: thus, our search targeted papers published between January 1970 and August 2019.} 

With the above procedure we elicited an initial set of 609 papers, of which 159 were on software project failures and 434 on software project successes with the others describing case-studies or direct practitioner experience without any specific success or failure discussion. 

\revised{Then, we executed the same manual filtering process of M{\"a}ntyl{\"a} et al. \cite{MantylaJRE17} to remove non-relevant sources. Specifically, we filtered out:}

\begin{itemize}

	\item \revised{papers that were not written in English;}
    
	\item \revised{papers whose full text was not available;}
    
    \item \revised{short papers (up to 4 pages) just reporting preliminary results;}
    
    \item \revised{papers from workshops;}
    
	\item \revised{papers that adopted the term ``failure" to indicate software faults;}
	
	\item \revised{papers that described a method or tool that theoretically would reduce the risk of project failure or increase the likelihood of success, but that did not actually assess it;} 
    
    \item \revised{papers that described the failure or success of introducing new tools or processes, but that did not relate this to project success or failure;}
    
    \item \revised{papers referring to just one  development phase rather than the entire software lifecycle;}
    
    \item \revised{papers that were about project success and failure, but that did not provide research results;}
    
    \item \revised{duplicate papers; specifically we excluded conference papers in case an extended journal article version was available.}
    
\end{itemize}

\revised{The manual filtering was conducted by two of the authors of this paper, who jointly scanned each candidate paper and judged  its suitability for the study. This initial process took two weeks and led to the final selection of 89 primary studies,  almost evenly balanced between success and failure. These papers all come from well-established and high-ranking\footnote{either class A* or A or B from the CORE Edu Rankings portal \url{http://portal.core.edu.au/}} conferences and journals sponsored by ACM, IEEE, Springer, Elsevier, and Wiley.}

\subsection{Analysis and Synthesis Methods}\label{sec:methods}

\subsubsection{Qualitative Analysis}\label{gtan}
Analysis and synthesis of results were carried out through the well-known Grounded-Theory (GT) approach \cite{gt}. 

\begin{figure}
\begin{center}
\includegraphics[scale=0.5]{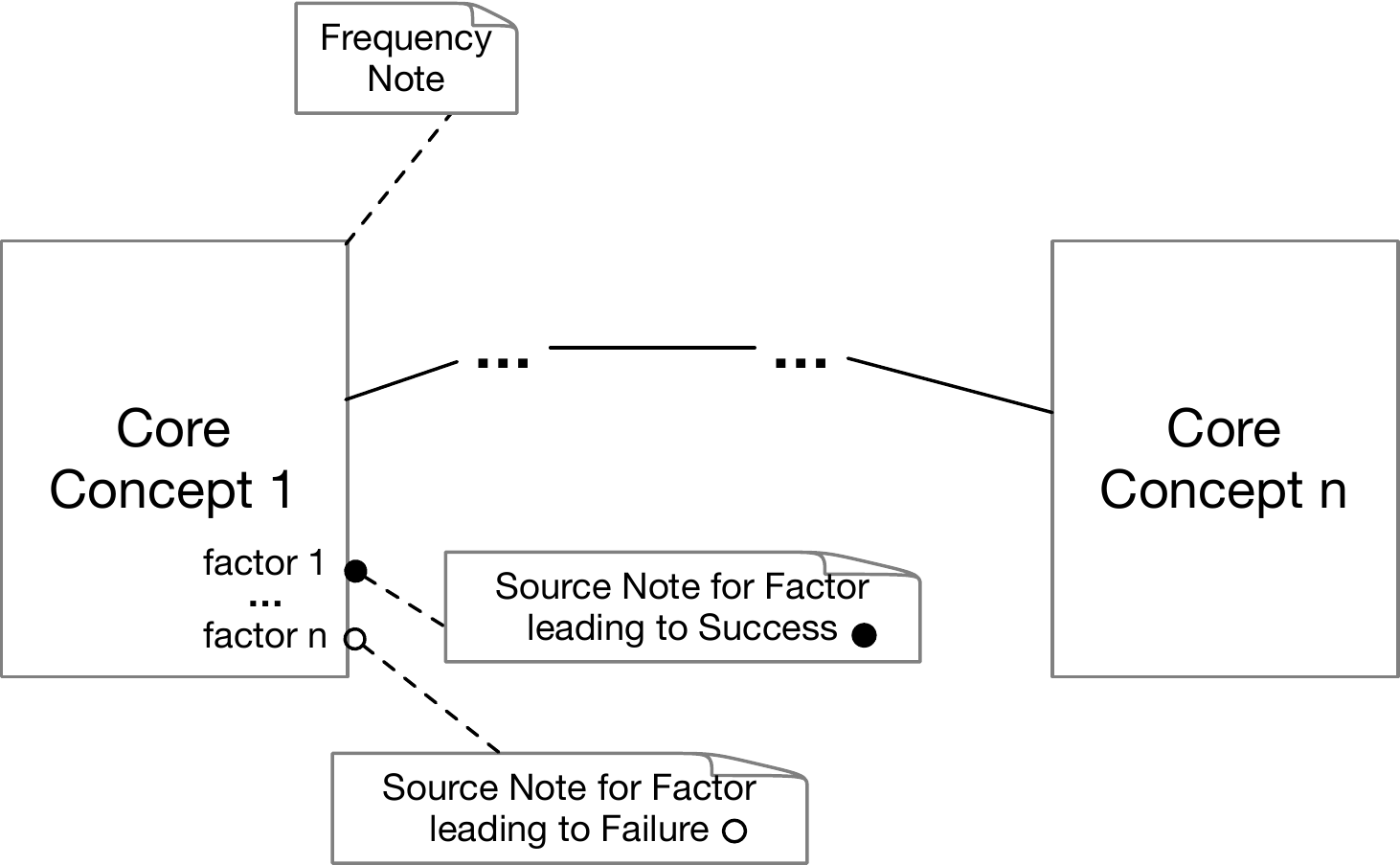}
\end{center}
\caption{Grounded Theory as Employed in This Study.}\label{ccrep}
\end{figure}

Fig. \ref{ccrep} outlines how we realized and represented the theory. Specifically, core concepts are represented as boxes with factors as attributes; relations reflect either memos or explicit relations found between factors. Every cluster is mapped to a note reporting its frequency and relative weight (measured in terms of reported code occurrences, and in how many papers those codes were reported) while every factor is mapped to a reference literature element with indication of whether the factor was leading to success (filled circle) or failure (unfilled circle). \revised{For example, the usage of the buddy-pairing best practice as part of Cisco systems' strategies to address global software engineering from one of our success-story reports is tagged with the ``best practice" code, as well as the ``global-software engineering process" code}.

\subsubsection{Quantitative Analysis}
Following an  approach similar to that proposed by M{\"a}ntyl{\"a} et al. \cite{MantylaJRE17}, we used the well-established topic modelling technique known as Latent Dirichlet Allocation (LDA).  However, rather than applying the technique to our papers as done previously, we applied the technique to cluster the factors emerging from our grounded theory activities, along with their textual definitions. Clustering of such factors allowed us to elicit a detailed view of the factors themselves, thus enabling the extraction of valuable themes among them. Furthermore, to preserve the relations elicited through grounded theory, the cluster analysis was conducted using the native XMI formatted files extracted from the models defined previously in Sec. \ref{gtan}.

%Much like our target study, in the scope of the clustering exercise, we performed standard text-mining pre-processing aimed at improving results by removing unnecessary information. Specifically: (1) all terms and definitions for the factors were standardised in terms of structure (i.e., definition + sample text extracted from reference papers); (2) punctuation marks and numbers were  removed; (3) all letters were converted to lower case; (4) all common stop words for English grammar and syntax were  removed.
%; (5) stemmed the text under analysis, created a document-term-matrix, and removed terms with scores below the median term frequency minus inverse document frequency (tf–idf) from our vocabulary.

For this topic modelling exercise, log-likelihood was used to assess clustering appropriateness. We began with the same number of clusters as the target study (k = 10 clusters) but that number was increased until at least one of the newly-emerging clusters contained less than half of the mean population of factors in the previous round. This approach was aimed at allowing the extraction of themes that were meaningful, i.e., they reflected semantic commonalities among factors. In addition, We used hyperparameter tuning over LDA hyperparameters alpha and beta \cite{AgrawalFM16}. To conduct all the above pre-processing and analyses we exploited the NetCulator bibliometric analytics tool\footnote{\url{https://www.netculator.com/}} which supports LDA and a number of similar natural-language analyses and clustering techniques.

\section{Results}\label{res}
\subsection{A Grounded Theory of Success and Failure: General Overview}
\label{sec-gt}

The entire grounded theory we elicited cannot be trivially represented and reported here because of its size and extensive detail. The grounded theory counts 561 factors and 40 core concepts in total, linked by 84 co-occurrence relations. However an overview is available to browse  as an online image.\footnote{\url{https://tinyurl.com/y79hfvby}} Furthermore, the grounded theory is available for further study as part of our replication package in three formats: MagicDraw resident UML format, XMI 2.11, and PDF. 

\begin{figure}
\begin{center}
\includegraphics[scale=0.4]{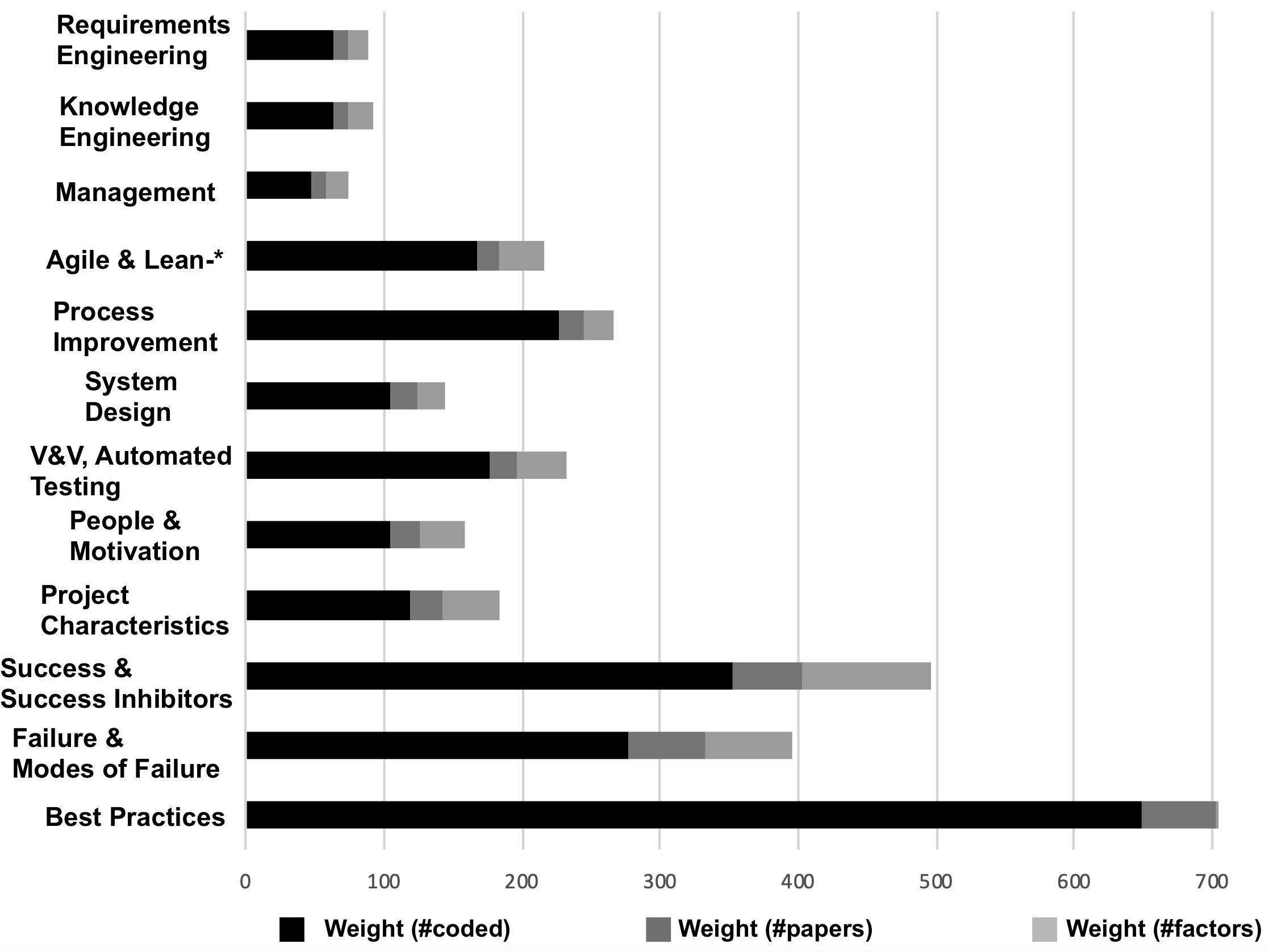}
\end{center}
\caption{Grounded Theory Content Analysis.} \label{overviewplot}
\end{figure}

In the scope of this study and to address \textbf{RQ$_1$} we offer an outline of the core-concepts and their content analysis \cite{HsiSha05}. Figure \ref{overviewplot} plots an overview of the most frequently occurring core-concepts captured with the method as described in Sec. \ref{gtan}. This plot shows the clusters ordered from top to bottom by increasing number of coded papers per cluster; every bar reports the stacked numbers of (1) coded occurrences, (2) papers in which the codes were applied for the core-concept, and (3) number of factors reported for the core concept. Occurrences reported on 9 papers or fewer were omitted for the sake of readability. %While the full data overview is available in the provided replication package, we could observe that the major component influencing success and failure of software projects is the availability or lack of best practices, 

The clusters we report reflect an equal mix of typical software lifecycle phases (e.g., \emph{requirements engineering}, at the top of Fig. \ref{overviewplot}) as well as practices used in those phases (e.g., \emph{V\&V and Automated Testing}). \revised{Moreover, the clusters reflect varied levels of abstraction among core concepts. Remarkably, the most frequent code is in the low-abstraction spectrum of the aforementioned level; according to our analysis the application of best practices as well as their success appraisal (bottom of the plot) is the most frequent code cluster. This evidence reflects that best practices as a construct of software engineering shows a presence which is comparable to the most frequent core-concepts. This indicates, (1) a gap in the levels of abstraction concerning both success and failure as evident from the state of the art and (2) a relative distance in the depth of knowledge in the respective clusters found.} Specifically, the clusters reflect the definitions outlined below:

\begin{enumerate}
\item \revised{\emph{Requirements Engineering.} Factors in this cluster address the creation, processing, resolution, traceability or quality of requirements as well as any factors influencing any phases of their lifecycle. Example factors include \emph{requirements validation by end users} as well as \emph{use of adequate language with the stakeholders}.}

\item \revised{\emph{[Software] Knowledge Engineering.} Factors in this cluster address the creation and retrieval of knowledge and artifacts of a software design, its implementation, and its operations. Sample factors include \emph{tacit knowledge} as well as \emph{knowledge brokers}.}

\item \revised{\emph{Project Management.} Factors in this cluster address the fallacies and pitfalls manifesting during, or related to the role of project management. Factors include the \emph{choice of software development model} as well as \emph{post-mortem analysis}.}

\item \revised{\emph{Agile and Lean-*.} Factors in this cluster address any positive or negative characteristics agile tenets, according to the definitions in Schwaber \cite{Schwaber04} and Kumar \cite{kumarlsd}. Sample factors include  developer software production \emph{worflow awareness} as well as \emph{human agile metrics}.}

\item \revised{{Process Improvement.} Factors in this cluster refers to the quantities and qualities of software processes that can be tracked and measurably improved. For example, the \emph{adoption of a common vocabulary} or the development of a shared \emph{vision}.}

\item \revised{\emph{System Design.} Factors in this cluster refer to the pros, fallacies, and pitfalls surrounding or in connection to a system's design and designers. For example, factors include \emph{software design reviews} as well as \emph{detailed design verification}.}

\item \revised{\emph{Verification \& Validation, Automated Testing.} Factors in this cluster refers to connections between  software success/failure and its V\&V processes and tools. Sample factors include \emph{design-for-testability} and \emph{ensuring test coverage}.}

\item \revised{\emph{People \& Motivation.} Factors in this cluster address the human and organizational issues of people and their motivations. Sample factors include \emph{low staff turnover} and \emph{supportive relationships}.}

\item \revised{\emph{Project Characteristics.} Factors in this cluster cover characteristics of projects and their role as mediators for success and failure. Sample factors include \emph{project lifespan} and \emph{object-level concurrency control}.}

\item \revised{\emph{Success and Success inhibitors.} This cluster contains success drivers and factors contributing to its definition and its inhibition. Such factors include \emph{errors in tracking the actual costs and debts} as well as \emph{dependency on other projects}.}

\item \revised{\emph{Failure and modes of failure.} This cluster contains failure drivers and factors contributing to software failure as well as modes of failure reported in literature. Factors include \emph{failure reticence} and \emph{wrong automation of manual processes}.}

\item \revised{\emph{Best-Practices.} This cluster contains factors relating to the definition, application, and successful appraisal of \emph{best practices}. Out of all the clusters, this one the least abstract, containing just three sub-factors: (1) \emph{the degree of dissemination and use of best practices}; (2) \emph{the appraisal of successful use of best practices}; (3) \emph{preconditions for best practice use}.}

\end{enumerate}

\subsection{Success and Failure Distilled: Topic Modelling Results}\label{tmod}

\begin{figure}[H]
\begin{center}
\includegraphics[scale=0.6]{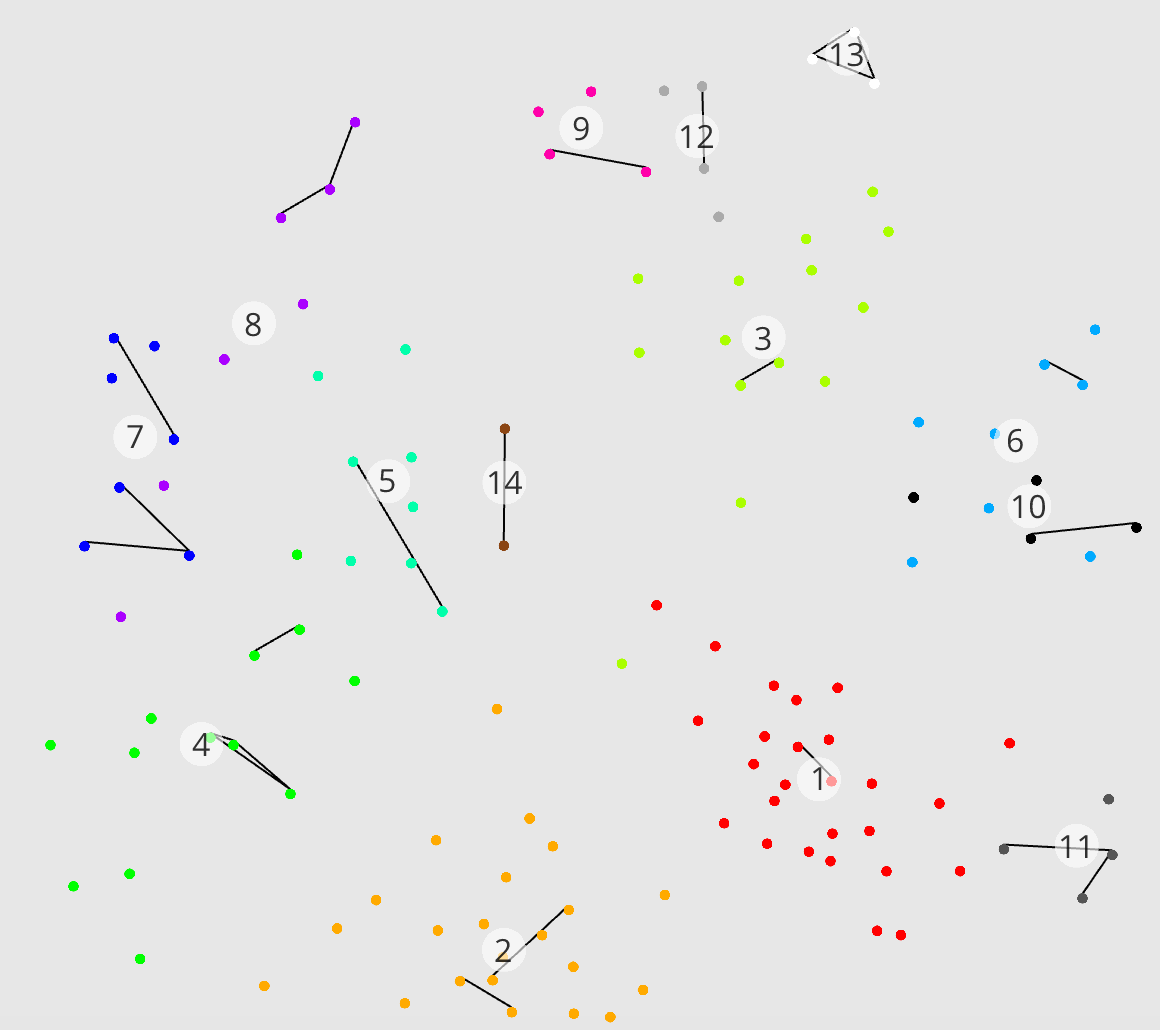}
\end{center}
\caption{Topic Modelling Clustering topology analysis \cite{topology} results.}\label{clusters}
\end{figure}

\begin{figure}
\begin{center}
\includegraphics[scale=0.42]{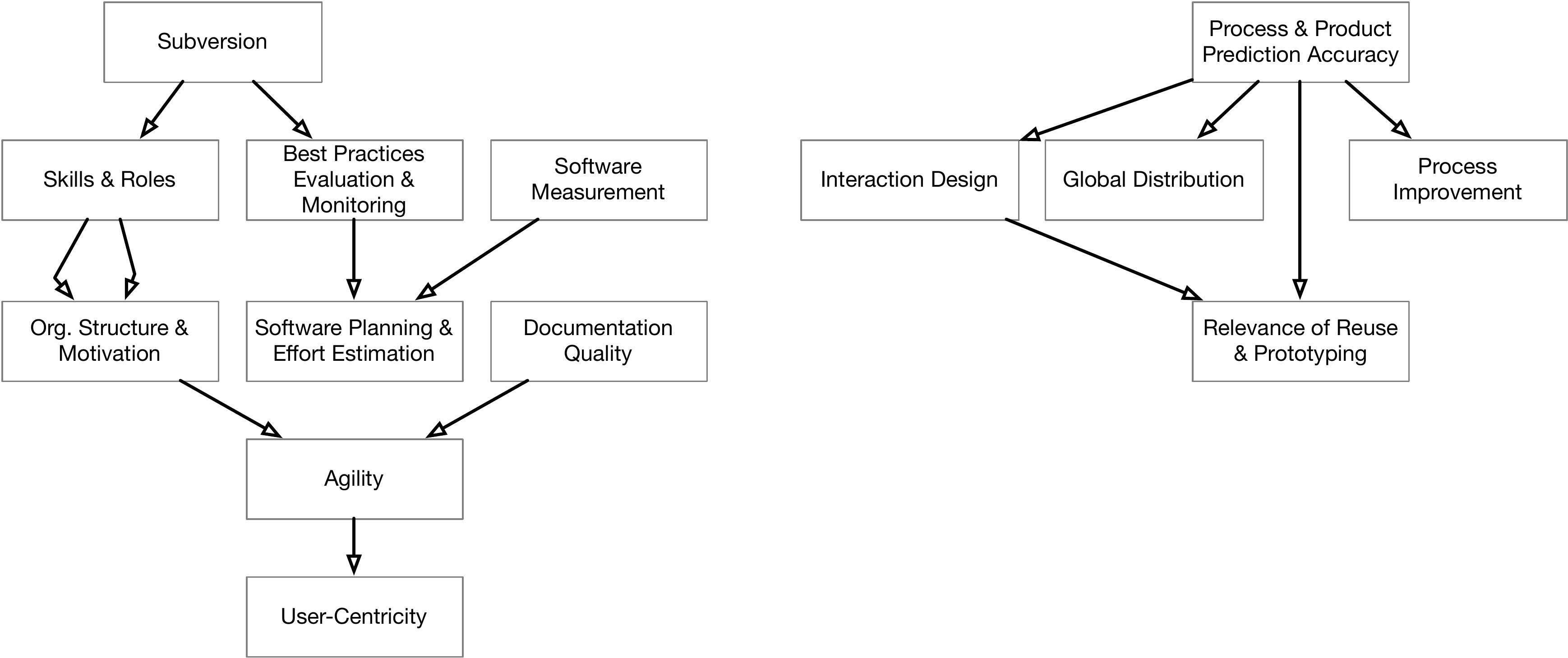}
\end{center}
\caption{Summary of Topic Modelling Results.}\label{topics}
\end{figure}

The results of the topic modelling exercise are recapped in Fig. \ref{clusters} and Fig. \ref{topics}. 

Fig. \ref{clusters} outlines the raw results of the 14 themes emerging from topic modelling. Dots indicate most-probable word radixes belonging to each theme (not reported on the figure for the sake of brevity but highlighted later in this section); edges reflect relations among core concepts from our grounded theory exercise.

On the other hand, figure \ref{topics} reports a manual representation of the themes emerging from the topic modelling exercise. Names for the themes were chosen independently by two analysts, with subsequent conflict resolution ($K_{alpha}=0.89$). For the manual creation of the figure we also used the relations previously reported as part of the grounded theory exercise (directed arrows in Fig. \ref{topics}). However, for the sake of visualization, all relations were collapsed into a single (unweighted) arrow linking the clusters occurring in each relation. Based on the relations, the emerging sets of themes self-arranged into two domain areas that delimit the phenomena under study (success and failure of software engineering projects). 
 
The area on the left-hand side of Fig. \ref{topics} incorporates \emph{people} themes (subversion \cite{RostG09}, organizational structure and motivation \cite{ossslr}, agility \cite{Abbas09}) as well as themes that discuss \emph{internal software product characteristics}, that is, themes of characteristics of the product which are not perceived externally by end-users---specifically, the quality of documentation \cite{garousi2013evaluating}, user-centric design \cite{SoyluCPBD11}, software measurement \cite{Fenton94}, software planning and estimation \cite{nasir_survey_2006}. Finally, this area contains  \emph{best-practices evaluation and monitoring} which is often considered orthogonal to all of the above but empirically is linked to the emergence of subversion \cite{dynos} and is reportedly connected to software planning and effort estimations.

The area on the right-hand side of Fig. \ref{topics} incorporates \emph{process-specific} themes (process improvement \cite{OgasawaraKA14}, accuracy of automated quality predictions \cite{AssarBP16}) as well as \emph{external product} themes, that is, characteristics of the software product that are perceived externally---specifically, the distribution of its software process \cite{jebertKP16}, its interaction design characteristics \cite{BellinghamHM14} as well as the extent to which external products have contributed to that product through prototyping and reuse \cite{Rubin97}.

The themes emerging in both domain areas are fleshed out in the following subsections, arranged left-to-right, and top-to-bottom following the contents of Fig. \ref{topics}. We provide definitions of themes and, as is typical for LDA-based topic modelling, we offer the list of the most important terms as determined by the algorithm (arranged by decreasing rank with a cut-off below 20\% probability) for each theme.

\subsubsection{Domain Area 1: People, Internal Software Characteristics, Best Practices}

\begin{enumerate}
\item \textbf{Subversion.} The concept of subversion refers  to concepts and challenges of subversize stakeholders previously introduced by Ross and Glass \cite{RostG09}. In the scope of our topic modelling exercise, this theme corresponds to most recurrent words and substrings (using wildcards) specified as follows: \emph{friction*, restrict*, *communication*, *cooperation*, disregardof*, lackof*};

\item \textbf{Skills \& Roles.} The concept of software skills and appropriate role management is still under investigation from several perspectives \cite{Pham16,YangKM08}, though most prominently from an educational viewpoint. Concerning the theme, most recurrent words and substrings we reported are: \emph{soft*, *motivation, coaching, *experience, domain*, trust, *size,  core-*, connected*};

\item \textbf{Best Practices Evaluation \& Monitoring.} We reported a strong presence of factors and themes relating to the application, appraisal, and effectiveness measurement of best practices, intended as \emph{recurrent solutions for known and established problems} \cite{ebert2007practices}.  Most recurrent words and substrings for this topics are: \emph{defect-*, accura*, change-*, *prediction, *accuracy, *success, *practice};

\item \textbf{Software Measurement.} Software measurement \cite{Fenton94} is a key activity in the scope of software engineering research. Most recurrent words and substrings concerning this theme within the scope of our results are as follows: *quality, define*, instrument*, customer*, stakeholder*, *community*, *interpretation;

\item \textbf{Org. Structure \& Motivation.} Organizational structure refers to the graph of recurrent, explicit or implicit relations of coordination, co-operation, and communication relations occurring among individuals in an endeavor \cite{ossslr}. The terms occurring for this theme reflect a prominent role of motivation as a driving force. Specifically, recurrent words and substrings are: \emph{turnover, *motivated, environment, feedback, recognition*, *motivator*};

\item \textbf{Software Planning \& Effort Estimation.} A well-established area of software engineering research and practice, software planning and effort estimation are key activities in software engineering economics \cite{Boehm1981}. In the scope of our work, most recurrent words and substrings relating to this theme are: \emph{misuse, earn*, staff*, governance};

\item \textbf{Documentation Quality.} From the perspective of software maintenance \& evolution, documentation is a discriminant in successful or failing software projects \cite{Boyd04b,Borja2000,1085331}. We obtained the following recurrent words and substrings for this theme: \emph{*knowledge*, domain*, requirements*, formal*, granularity, broker*, post-mortem};

\item \textbf{Agility.} Agility clearly relates to the use, level of, and confidence around agile methods \cite{agilesmells}. The adoption of agile methods is an established fact in software engineering literature \cite{noKey}; however, the factors that lead to successful or failing attempts at harnessing agile methods are still left largely to speculation. In the scope of our topic modelling exercise, the following terms were reported: \emph{self*, user*, value*, pressure*, pair*, test*, human*};

\item \textbf{User-Centric Design.} Finally, in the scope of topics relating to people, internal software characteristics, as well as best practices, we reported several factors and recurrent keywords relating to user-centric design \cite{Rubin1994hou}, that is, the framework of engineering where usability goals, user characteristics, environment and workflows are given attention at each stage of the (software) design process. Many of the keywords reported for this theme relate to how practices from this framework lead to successful or failing engineering attempts. Specifically, words and substrings reported are: \emph{persona*, communit*, organization*, usabilit*, integrat*, context*};
\end{enumerate}

\subsubsection{Domain Area 2: Processes and External Product Characteristics}

\begin{enumerate}
\item\textbf{Process \& Product Quality Prediction Accuracy.} This theme relates to the accuracy with which a quality prediction is made or appraised in the scope of software engineering research \cite{Jorgensen1995,ChenH09}. Several works from the literature have touched upon this topic, most prominently along the lines of defect prediction \cite{AssarBP16} and similar endeavours. Words and substrings featured in this theme are: \emph{histor*, objective*, improvement, additional*, technolog*};

\item\textbf{Interaction Design.} Interaction design refers to the design of interactive products and services in which design focus goes beyond the product under development and includes the ways users are likely to interact with that product \cite{0470084111}. Although not a common software engineering topic of focus, interaction design reflects several keywords occurring frequently in general software engineering literature, most prominently: \emph{socio-*, man-machine*, cognitive*, anthropo*, bond*, operation*};

\item\textbf{Global Distribution.} Global distibrution in the scope of the themes emerging from our topic modelling refers to the general sub-field of software engineering that studies globally-dispersed teams as part of global software engineering and development \cite{gse,icgseoss}. The most frequent words and substrings relating to this theme are: \emph{remote*, geograph*, standard*, expan*, distribut*, multi*, organization*};

\item\textbf{Process Improvement.} Process improvement refers to the segment of software engineering research and practice dedicated to appraising and improving the quality of software processes \cite{Grady1997,niaym2006}. In the scope of our topic modelling, words and substrings relating to process improvement are: \emph{progress*, train*, ad-hoc, capabilit*, principl*, chang*, need*, expectation*, assess*};

\item\textbf{Reuse \& Prototyping.} The last emerging theme out of topic modelling reflects the role of software reuse and rapid prototyping as strategies for software engineering, where \emph{reuse} indicates the recycling of existing software assets into a new or evolved version of a software product \cite{Frakes1994} while \emph{prototyping} reflects the preparation of mock-ups for exploratory requirements engineering \cite{ThompsonW92}. Key terms for this theme are: \emph{decreas*, upgrad*, reverse, cost*}.
\end{enumerate}

%\subsection{Community Smells}

\section{Usage, Implications and Threats to Validity}\label{disc}
\subsection{Discussion}

\revised{Our results indicate that the phenomena of software success and failure are extensive and span a large variety of factors and themes, not all of which are currently measured or tracked. Furthermore, there seems to be a mismatch or some form of \emph{failure reticence} in the field, since the literature reports a majority of studies focused on software success as opposed to failure.}

\revised{We conclude that further research should be dedicated into both the phenomena under study, but emphasize that such research should elaborate more on the phenomena associated with software failure, the factors entailed, and their many relations and ramifications.}

%Our results build upon the contributions and conclusions of the previous replication target. Specifically, the study in question concludesd with 3 key challenges. 
\revised{Stemming from previous studies, we renew the conclusions of those studies with our own data and observations. In addition, we provide three other observations:}
\begin{enumerate}
\item \emph{Creating and Validating Instruments for Measuring Success} \revised{--- we confirm this finding from multiple perspectives.  For example, we discovered that the correct use and appraisal of best practices in software engineering is least understood and yet such understanding is urgent since it often mediates software failure and success altogether.}

\item \emph{Representative Sampling Without Population Lists} --- \revised{although we did not conduct any specific analysis to confirm this finding, we did in fact report a relative paucity of methodological detail in about 70\% of the papers that we surveyed. The lack of rigour and replicability compromises the generalisability of individual findings\revised{\footnote{Nevertheless, to encourage replication, a comprehensive replication package is provided online here: \url{https://figshare.com/s/e6f0968e55c2cd024389}}}.}

\item \emph{Identifying Empirically Validated and Actionable Antecendents} --- \revised{similarly to the previous point, we did not conduct any systematic analysis focusing on the antecedents in question but we did report a relative lack of dimensions, factors, and valid metrics from a considerable subset of the primary studies. Specifically about 60\% of the primary studies do not conclude with measurable quantities to be tracked and improved. }
\end{enumerate}

\revised{Furthermore, the study highlights several other findings, most prominently on the importance of the dimensions of \emph{subversion} around software, described in both this study and its precedent as the process whereby the values and principles of an established software engineering project are undermined, in an attempt to transform the social order and its structures of power, authority, hierarchy, and social norms in line with some desired end-state differing from the project goal. Our findings highlight a prominence of \emph{subversive} dimensions. The existence and prominence of such dimensions further motivates streams of inquiry around social software engineering \cite{sosoen} and the quality of organisational and community structures  \cite{Palomba2018,tamburri2018discovering,TamburriKF16} for software engineering.}

\subsection{Addressing the Research Questions}
This study set out to address three research questions, namely: (1) What factors are reportedly connected to success or failure? (2) What themes emerge across such factors? And (3) What measurable quantities exist in  themes that are not currently being measured?
In addressing these research questions we reported, in the scope of \textbf{RQ$_1$}, the following:

\begin{framed}
\textbf{Answer to RQ$_1$.} There exist over 500 factors arranged in over 40 topical clusters of factors. Among these clusters, the most impactful in terms of occurrence and frequency (established via content analysis) range from software engineering phases such as \emph{requirements engineering} to the use and effectiveness-appraisal of \emph{best practices}. Further research can use the isolated clusters (and the factors therein) to devise tools and metrics for continuous monitoring and analysis.
\end{framed}

Furthermore, in the scope of \textbf{RQ$_2$}, we aimed to determine  additional themes within the factors, beyond those found in our manual qualitative clustering. For this second endeavour, we reported the following:

\begin{framed}
\textbf{Answer to RQ$_2$.} There exist 14 underlying themes  among the over 500 factors in our analysis. Themes emerging from this analysis constitute essential risk engineering targets for successful software engineering.
\end{framed}

Based on our results and the answers to both research questions, the two perspectives that may make practical use of the synthesis that we have provided reflect (1) practitioners' efforts in avoiding failure and (2) researchers' efforts in figuring out and measuring both success and failure.

On one hand, practitioners can focus on the factors (and clusters thereof, see Fig. \ref{overviewplot}) that reflect (1) \emph{success and success inhibitors}, (2) \emph{failure and failure modes} as well as (3) \emph{best practices} and their evaluation. In so doing, practitioners can use the factors we provide as indicators to assess their project status and can plan and instrument corrective actions.

On the other hand, researchers can use the theoretical modelling exercise reported in Fig. \ref{overviewplot} to further understand and potentially measure the factors, focusing on operationalising any factors that were not previously measured. At the same time, the topic modelling exercise we reported in Sec. \ref{tmod} could be used as a basis to design, prototype, and evaluate automated computational intelligence \cite{2008-115} methods, tools, and techniques to automatically determine the status of software projects, e.g., analyzing data stemming from the DevOps pipelines around such projects.\footnote{\url{https://dzone.com/articles/role-of-predictive-analytics-in-devops}}

Finally, in the scope of \textbf{RQ$_3$}, we set out to identify the dimensions emerging from the previous analyses which, to date, do not have any automated means of measurement, tracking, and improvement in software engineering research and practice. To address this gap, we elaborated a quality model \cite{InternationalStandardOrganization(ISO)2001} obtained by identifying the factors from our study (\textbf{RQ$_1$} and \textbf{RQ$_2$}) which are currently \emph{not} supported by any  artefact corresponding to the definition of a quality model \cite{Dromey95}. A quality model establishes relationships between project quality outcomes (e.g., bug rates, issue resolution time, size and vigor of the community, etc.) and characteristics of the product and its community. The next section outlines this contribution in more detail.

\subsection{A Quality Model for Unobserved Software Quality Dimensions} 

To address the gap identified by \textbf{RQ$_3$} we operated a simple systematic search of every keyword discovered as part of topic modelling (see \textbf{RQ$_2$}, Sec. \ref{res}) along with the additional search string defined as follows:

\begin{center}
quality $\wedge$ (model $V$ framework $V$ metric $V$ measure $V$ measurement $V$ analysis $V$ parameter);
\end{center}

As a result of this exercise, our model addresses 3 unobserved themes: (1) subversion; (2) organizational structure and motivation; (3) skills and roles.

We aggregated all metrics and empirically-investigated quantities from software engineering research that emerged from the systematic search above. The metrics and quantities involved are all related to features and characteristics of a social graph construct, known as \emph{Developer Social Network}, loosely defined by Meneely and Williams \cite{Meneely2011} as the superimposed communication and collaboration networks structures emerging during software development. The aforementioned construct was previously touched upon by several other research attempts, also in relation to software failure \cite{pinzger2008cds}. We reuse this construct as a reference to flesh out the metrics we discovered in literature that address the aforementioned observation gaps. A total of 38 metrics were found.

An elaboration in full detail of all the 38 metrics for of each quality category featured in the model is outside of the scope of this contribution, which is aimed at offering an aggregate quality model rather than a detailed treatise or synthesis of each factor.\footnote{For complete details, the reader may refer to \cite{thesis} which contains a complete overview of all factors in the quality model, their operationalisation, and their implementations in practice.}
%The interested reader is referred to the reference literature linked in the full technical report \cite{} with the entire detail of the quality model.

The emerging quality model features 5 categories of previously-defined,  validated metrics that can aid the observability of \emph{subversion}, \emph{organizational structures \& motivation}, as well as \emph{skills \& roles}. These metrics span: (1) developer social networks (DSNs)  --- these mainly reflect population metrics applied in the context of DSNs \cite{AmritSASNA}; (2) socio-technical --- these mainly reflect quantities that were introduced to relate communication (i.e., information interchange) and collaboration (i.e., co-operated action over software artefacts) together, most prominently socio-technical congruence \cite{congruence}; (3) core-community members --- these mainly reflect  the difference between features in the \emph{core} and \emph{periphery} of the network structure \cite{CrowstonWLH06,AmritH10}; (4) turnover --- these mainly reflect  the degrees of freedom or variability of members within the DSN; (5) social networks analysis (SNA) --- these mainly reflect the use of ``classical" SNA metrics that were previously applied in the context of software engineering \cite{AllahoL15}. To address \textbf{RQ$_3$} we  argue as follows:

\begin{framed}
\textbf{Summary for RQ$_3$.} There exist three themes emerging from our systematic literature analysis that are currently not supported by a full-fledged quality model. They are: (1) subversion; (2) organizational structure and motivation; (3) skills and roles. Nevertheless, there exist in the literature a considerable number of metrics to address the aforementioned gaps. These metrics are openly available online~\cite{thesis} and reflect 5 categories of quality that need to be explicitly tracked to monitor the extent of software success and to ward off software failure. The proposed quality model can be used in conjunction with established technical, process or other quality models for software engineering practice.
\end{framed}

\subsection{Observations and Implications}

First, from a purely statistical perspective, the clusters and themes discussing {\em best practices}---their evaluation and monitoring as well as software success and failure---were the most popular ones emerging from this study. Furthermore, these themes and clusters emerged \emph{both} from (1) topic modelling and (2) grounded theory. And this topic by far outweighed all others in terms of software engineering research and practice. This finding confirms what was previously reported in M{\"a}ntyl{\"a} et al. \cite{MantylaJRE17}. \emph{Further research should thus be dedicated to establishing this research cluster/theme as a research topic in its own right.}

Second, based on the extent of our data (500+ factors over 40+ clusters), software success and failure are vast phenomena which deserve dedicated  software engineering research on their own. Specifically, the dimensions and factors along which success (or failure) unfold need statistically significant factor analysis using time-series analysis \cite{CoutoMGV13} or similar approaches to effectively establish what factors and dimensions  contribute to or facilitate success. Conversely, our data indicates that we know much more about success than we do about failure (e.g., see Tab. \ref{overviewplot}). The number of codes applied for the core concepts of \emph{success} and \emph{failure} differ by almost 2 to 1 and the number of papers in which these codes were applied is 1.7 times higher for \emph{success}. To address this gap like other engineering disciplines, software engineering research should dedicate research to establish more background knowledge on software failure (e.g., reflecting post-mortem analysis \cite{Hager91}, empirical software failure research, fault lines \cite{BahmaniSS018}, etc.). \emph{In summary, further research along this line should be dedicated to better understanding software failure, perhaps starting from well-known cases of software failure, e.g., in open-source. Specifically, open-source phenomena such as forge failure, community forking, and sustainability beyond forks are still not widely studied and thus deserve further empirical and experimental research.}

Finally, our 3 \textbf{RQ}s together amount to a single key message: \emph{software engineering is a perilous game of equilibrium over as many as 500+ degrees of freedom. Constant feedback loops between all areas of the organisational and technical structures involved, be they open- or closed-source, is required to maintain this equilibrium. Sustaining these feedback loops by any means necessary should be a key goal for future software engineering research.}

\subsection{\revised{Threats to Validity}}

\revised{The conclusion provided by our study might have been threatened by a two main factors: the collection of a complete set of papers on the subject of interest and the way we analyzed the collected sources to provide new knowledge.} 

\revised{In the first instance, the major challenge of a systematic literature review is that of finding a comprehensive set of papers to study and analyze. In our case, we built a search string that not only included keywords coming from the reference work of M{\"a}ntyl{\"a} et al. \cite{MantylaJRE17}, but also aimed at retrieving papers offering results stemming from industrial practice and experience. Using this strategy, we were able to survey the literature on success and failure more comprehensively and from different perspectives. In so doing, we queried all major databases currently available in software engineering research, hence increasing the comprehensiveness of our research.  Furthermore, it is worth noting that two authors jointly scanned each of the papers coming from the application of the search string with the aim of (1) assessing its fitting to the goals of the paper, thus discarding non-relevant ones by means of the exclusion criteria defined in Section \ref{sec:literatureMethod} and (2) increasing the overall reliability of the methodological procedure, by conducting a joint effort in evaluating it.}
	
\revised{When analyzing the sources retrieved after the application of the search string, we applied formal grounded theory methods to let emerge themes related to software engineering success and failures. To increase the reliability of the applications of such a methodology in our context, two authors of this paper have jointly performed the task: they analyzed each of the retrieved sources to understand concepts and assign codes. Furthermore, to ensure internal and construct validity even further, the set of codes for grounded theory was later double-checked by an external researcher having more than 10 years of research experience, who fully confirmed the initial codes assigned by the two authors of this paper. With these steps, we aimed at increasing the overall validity and reliability of the reported results; nevertheless, we cannot exclude possible imprecision and/or subjective judgment that may have played a role in the elaboration of the codes. For these reasons, we make our data publicly available to enable further replications and verification of our analyses.}

\section{Conclusions}\label{conc}
\revised{This section reports on the practical usage of the results achieved in our study and outlines our future research agenda on the topic.}

\subsection{\revised{Results Usage in Practice}}
\revised{From a more practical perspective, the results provided in the previous pages can be used in at least four practical scenarios.}

\revised{First, practitioners steering their own software engineering endeavours can use the overview provided in Fig. \ref{overviewplot} and \ref{topics} to understand the potential areas at risk within their software projects. Later, once these areas are understood, practitioners can use the more fine-grained and detailed grounded theory to pick and choose which factors are known inside those sensitivity areas. In the same vein, practitioners can also bootstrap new software engineering endeavours providing an appropriate software risk analysis starting from the results we have provided.}

\revised{Second, practitioners can use the metrics and indicators \revised{accounted for in our grounded-theory or any of its syntheses in this manuscript as input for organizational quality tracking and continuous improvement,} just as technical metrics are used to track and improve software coding practices. In line with this contribution, we have designed and implemented a research tool to automate the elicitation and analysis of such metrics. This tool is being refined based on a fork of the Siemens CodeFace tool~\footnote{\url{http://siemens.github.io/codeface/}} and is currently under experimentation.\footnote{The tool is available and free to use under the Apache 2 license agreement: \url{https://github.com/maelstromdat/CodeFace4Smells}}}

\revised{Third, practitioners and software vendors active in the quality assurance software tools market segment can use the factors and reference analyses in the scope of our \textbf{RQ}s to refine their tools in line with the findings of this study or even devise new tools to support the unobservable dimensions isolated as part of our response to \textbf{RQ$_3$}.}

\revised{Fourth, practitioners can conduct a self-assessment of their software projects with respect to the factors we summarized in the previous sections. A rudimentary risk self-assessment methodology entails at least the following steps:}

\begin{enumerate}
\item \revised{Download the grounded-theory model we have provided online;\footnote{\url{https://tinyurl.com/y79hfvby}}}
\item \revised{Use the model as a checklist to assess whether failure-inducing factors (those marked with an empty circle linked note reporting the papers discussing them) may be leading to risks of failure;}
\item \revised{Use the model as a checklist to assess whether success-facilitating factors (those marked with a filled circle linked note reporting the papers discussing them) are reflected in the project under study;}
\item \revised{Elaborate the total risk of failure as follows:}
\begin{enumerate}
\item \revised{\textbf{Elaborating the Known Risks.} Subtract the \emph{positive knowns}, that is, the sum of known success-facilitating factors exerting an observable effect on the project from the \emph{negative knowns} exerting an observable effect on the same project. This is reasonable since risk is higher if negative factors are manifested but can be lowered to the degree that positive and success-inducing factors are manifested;\footnote{This assumes that all factors have an equal mutual effect, which is obviously an open research question.}}
\item \revised{\textbf{Elaborating the Unknown Risks.} Sum together any remaining negative and positive \emph{unknowns} from the model.  This is reasonable since the risk of failure is higher the more factors' effects are unknown to an observer, regardless of whether those effects are positive or negative;}
\item \revised{\textbf{Elaborating a Grand Total.} Sum together the two compounding quantities above.}
\end{enumerate}
\end{enumerate}

\revised{The steps entailed in (4.a-c) allow practitioners to get a rough evaluation of the risk coverage for the project under study. More formally:}

\begin{center}
\revised{\emph{Software Failure Risk:\\}}
\revised{$\rho = \sum [(P_n - N_n) + U]$;}
\end{center}

\noindent \revised{where $P_n$ indicates the \emph{positive knowns}, while $N_n$ indicates the \emph{negative knowns} and U indicates any remaining \emph{unknowns}, e.g., accounting for contingency management and preparedeness planning. The above methodology and the basic formula are to be seen as a rudimentary starting point for further experimentation, which is beyond the scope of this study. However, we are planning several applications of the aforementioned methodology and formula in action in industry to elaborate more on its construct and external validity.}

\subsection{\revised{Synthesis and future work}}
\revised{This paper builds upon previous studies of the complex phenomena of software success and failure. The literature in question focuses on the software engineering domain and covers a broad range of perspectives over the discipline. In this paper we have presented a more extensive and rigorous analysis of the literature, by executing 3 analyses aimed at deepening our understanding of software success and failure. The 3 analyses reflect: (1) a grounded theory of the phenomena under study; (2) the emergent themes hidden beneath such a theory; (3) the measurable quantities from software engineering research that account for previously unobserved themes and factors from analyses (1) and (2) above.}
\revised{In the future we plan to further analyse the data and factors produced as part of our research question 1, e.g., to offer automated means of classification for the factors. Furthermore, we plan to analyse the data in our replication bundle for the purpose of generalising a more refined taxonomy or \emph{ontology} for the purpose of instrumenting automated reasoning and risk analysis (e.g., to support post-mortem analysis).}

\begin{figure}[H]
\begin{center}
\includegraphics[scale=0.3]{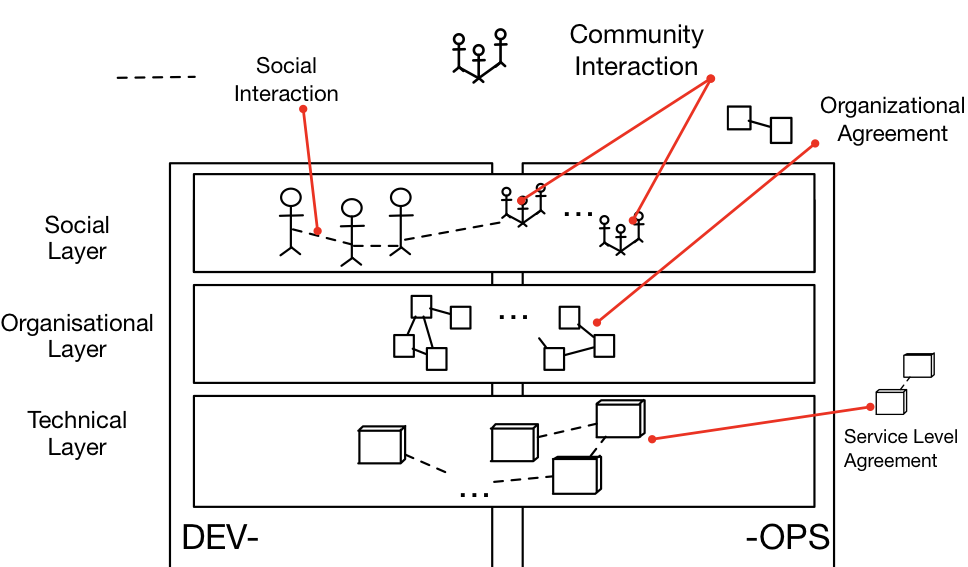}
\end{center}
\caption{\revised{Omniscient DevOps Analytics; concept tailored from \cite{tamburri2018omniscient}.}}\label{omniscient}
\end{figure}

\revised{Finally, we plan to refine and further evaluate tool support to track as many factors from our grounded-theory and themes as possible, automating their investigation from openly available application lifecycle management (ALM) tools of common use during software development, such as quality metrics suites, issue-tracking systems, CI/CD pipelines, and more. A vision for how this might be realized is presented in previous work \cite{tamburri2018omniscient}. Specifically, the unobserved themes emerging from this study could be supported for specifically-tailored holistic DataOps \cite{Ereth18} software process, product, and people analysis ALM suite \cite{KlespitzBK16} acting as an integrated predictive analytics solution working continuously towards modelling success and failure by means of machine-learning and similar advanced computational intelligence. In this vein, all the dimensions elaborated in our grounded-theory could be supported by specific predictive modelling computational intelligence while a holistic ALM suite back-end could be trained as an \emph{ensemble} method to assemble the individual predictions towards an aggregated series of fundamental scores, thus instructing all software stakeholders in their next steps. For example, see the recap in Fig. \ref{omniscient}; the figure outlines our future work towards a DevOps analytics suite which could be considered \emph{omniscient}, that is, acting towards most if not all of the dimensions accounted in the grounded-theory proposed in this work across all dimensions we highlighted, namely, the individuals dimension, their social-interactive communitarian dimension, the organisational layer combining them as well as the technical layer towards which their work is aimed.}

\section*{Acknowledgement}

The authors would like to thank Dr. Francesco Castri for his invaluable effort and work in producing and allowing for the inter-rater reliability assessment of the grounded theory reported in this paper. Furthermore, the authors would like to thank Dr. Massimiliano Di Penta, Dr. Henry Muccini, Dr. Patrizio Pelliccione, and Dr. Marco Di Memmo for their precious insights, explanatory, and confirmatory discussions in the scope of addressing the threats to validity of this work. Finally, some of the authors' work is partially supported by the European Commission grant no. 0421 (Interreg ICT), Werkinzicht, the European Commission grant no. 787061 (H2020), ANITA, European Commission grant no. 825040 (H2020), RADON, European Commission grant no. 825480 (H2020), SODALITE, and the Swiss National Science Foundation through the SNF Project No. PZ00P2\_186090 (TED). \revised{Finally, we would like to thank the associate editor and anonymous reviewers for the detailed and constructive comments on the preliminary version of this paper, which were instrumental to improving the quality of our work.}

\revised{
\section*{Appendix}
\begin{itemize}
    \item Downloadable Dataset:~\url{figshare.com/s/e6f0968e55c2cd024389};
    \item Primary studies\footnote{please consider that all copies are copyrighted and were downloaded for research purposes only.}:~\url{www.dropbox.com/s/3lj4h0otpt7jrxe/SLRWorkingPortfolio.pdf?dl=0};
    \item GT codes list and definition:~\url{www.dropbox.com/s/ia0yv225r8jcggr/code\%20list.xlsx?dl=0}
\end{itemize}
}

\bibliographystyle{unsrt}
\bibliography{surveynew}

\end{document}